\documentstyle[prl,aps,multicol,epsfig,amssymb,amsmath]{revtex}

\begin{document}

\def\ra{\rangle}
\def\la{\langle}
\def\ha{{\hat a}}
\def\hb{{\hat b}}
\def\hu{{\hat u}}
\def\hv{{\hat v}}
\def\hc{{\hat c}}
\def\hd{{\hat d}}
\def\no{\noindent}\def\non{\nonumber}
\def\hi{\hangindent=45pt}
\def\v{\vskip 12pt}

\draft

\title{Quantum Entanglement of Moving Bodies}

\author{Robert M. Gingrich and Christoph Adami}

\address{
Quantum Computing Technologies Group, Jet Propulsion Laboratory 126-347\\
 California Institute of Technology, Pasadena, CA~~91109-8099 }

\date{\today}

\maketitle

\begin{abstract}
We study the properties of quantum information and quantum entanglement in moving
frames, and show that the entanglement between the spins and the momenta of two
particles become mixed when viewed by a moving observer.  A pair of particles that
is entangled in spin but not momentum in one reference frame, may, in another
frame, be entangled in momentum at the expense of spin-entanglement. Similarly,
entanglement between momenta may be transferred to spin under a Lorentz
transformation.  While spin and momentum entanglement each is not Lorentz
invariant, the joint entanglement of the wave function is.
\end{abstract}


\begin{multicols}{2}

Mostly, theories in physics are born out of necessity, but not always. The
thermodynamics of moving bodies for example (relativistic thermodynamics) was a
hotly contested topic without resolution~\cite{old,ottarz6364,aldrovandi92,trout98}
(but see~\cite{landsberg96}) mostly because no experiment required it. As a
side-effect, it was learned that the temperature concept in relativistic
thermodynamics is ambiguous simply because radiation that is perfectly black-body
in an inertial frame is \textit{not} thermal if viewed from a moving
frame~\cite{peebles68,landsberg96}. This is an interesting result for information
theory~\cite{shannon48}, however, since if probability distributions can depend on
the inertial frame, then so can Shannon entropy and information. Even more
interesting are the consequences for quantum information theory, where quantum
entanglement plays the role of the primary resource in quantum computation and
communication~\cite{nielsenchuang00}. Relativistic quantum information theory may
become a necessary theory in the near future, with possible applications to quantum
teleportation~\cite{Bennettetal93} entanglement-enhanced
communication~\cite{BennettWiesner92}, quantum clock synchronization, and
quantum-enhanced global positioning~\cite{QClock}.

Entanglement is a property unique to quantum systems. Two systems (microscopic
particles or even macroscopic bodies~\cite{julsgaard2001}) are said to be quantum
entangled if they are described by a joint wave function that cannot be written as
a product of wave functions of each of the subsystems (or, for mixed states, if a
density matrix cannot be written as a weighted sum of product density matrices).
The subsystems can be said \textit{not} to have a state of their own, even though
they may be arbitrarily far apart. The entanglement produces correlations between
the subsystems that go beyond what is classically
possible~\cite{CerfAdami97}. It is this feature that enables quantum
communication protocols such as teleportation and super-dense coding. However, the
preparation, sharing, and purification of entanglement is usually a complicated and
expensive process that requires great care. It is therefore of some importance to
understand all those processes that might affect quantum entanglement (in
particular those processes that lead to decoherence). It was shown recently that
Lorentz boosts can affect the marginal entropy of a single quantum
spin~\cite{peresetal02}. Here, we determine that the entanglement between
\textit{two} systems depends on the frame in which this entanglement is measured.
We show that a fully entangled spin-1/2 system (a Bell state) loses entanglement if
observed by a Lorentz-boosted observer. Thus,
Lorentz boosts introduce a transfer of entanglement \textit{between} degrees of
freedom, that could be used for entanglement manipulation.  While the entanglement
between spin or momentum alone may change due to Lorentz boosts, the entanglement
of the \textit{entire} wave function (spin and momentum) is invariant.

In order to define the momentum eigenstates for a massive particle with spin, we
start by defining the rest frame eigenstates
\begin{eqnarray}
  P^\mu |{\bf 0} \lambda \ra      & = & |{\bf 0} \lambda \ra p_0^\mu \;,\\
  {\bf J}^2 |{\bf 0} \lambda \ra & = & |{\bf 0} \lambda \ra s(s + 1)\;,\\
  J_z |{\bf 0} \lambda \ra        & = & |{\bf 0} \lambda \ra \lambda\;,
\end{eqnarray}
where $p_0^\mu = (m,\bf{0})$, $s$ is the total angular momentum of the particle,
and $\lambda$ is the $z$ component of angular momentum.  Since the particle is at
rest, $s$ and $\lambda$ are the spin and the $z$ component of the spin for the
particle respectively.

We define a momentum state by acting on the rest frame state with a pure Lorentz
boost \begin{equation} |{\bf p} \lambda \ra \equiv L({\bf \xi}_{\bf p} ) |{\bf 0} \lambda
\ra\;. \end{equation} where $L({\bf \xi}_{\bf p})$ is a boost such that \begin{equation} L({\bf
\xi}_{\bf p}) (m,{\bf 0}) = (\sqrt{{\bf p}^2 +
  m^2},{\bf p})\;
\end{equation} where the rapidity, ${\bf \xi}_{\bf p}$, is given by
\begin{eqnarray}
  \sinh |{\bf \xi}_{\bf p}| & = & \frac{|{\bf p}|}{m} \\
  \frac{{\bf \xi}_{\bf p}}{|{\bf \xi}_{\bf p}|} & = & \frac{{\bf
  p}}{|{\bf p}|}\;.
\end{eqnarray}
In what follows, we use ${\bf p}$ to represent the 4-vector $(\sqrt{{\bf
p}^2 + m^2},\bf{p})$ unless it is ambiguous.

The effect of an arbitrary  Lorentz transformation $\Lambda$ (rotation and boost)
on a momentum eigenstate is
\begin{eqnarray} \label{wigner}
\Lambda |{\bf p} \lambda \ra & = & \Lambda L({\bf \xi}_{\bf p} ) |\bf{0} \lambda \ra\\
            & = & L({\bf \xi}_{\Lambda{\bf p}}) L({\bf \xi}_{\Lambda{\bf p}})^{-1}
            \Lambda L({\bf \xi}_{\bf p} ) |\bf{0} \lambda \ra\;.
\end{eqnarray}
Since $L({\bf \xi}_{\Lambda{\bf p}})^{-1} \Lambda L({\bf
 \xi}_{\bf p})$ leaves ${\bf 0}$ invariant it must be a
rotation.  These rotations are called the Wigner rotations $R(\Lambda,{\bf p})$,
and they act only on the rest frame spin component $\lambda$. Hence, we can write
\begin{equation} \label{lambdaonp} \Lambda {|{\bf p} \lambda \ra} = \sum_{\lambda'} |\Lambda
{\bf p} \lambda' \ra D^{(s)}_{\lambda',\lambda} (R(\Lambda,{\bf p}))\;, \end{equation}
where $D^{(s)}_{\lambda',\lambda} (R)$ is the spin $s$ representation of the
rotation $R$.  Here, we restrict ourselves to $s = 1/2$, but the generalization to
larger spins is straightforward. For a review of momentum eigenstates and spin,
see~\cite{tung}. Since a local unitary transformation will not affect any measure
of entanglement~\cite{vidal,parker}, the unitary transformation $\Lambda$ on the
infinite dimensional space of momentum and rest frame spin will not change the
entanglement between two particles provided we do not trace out a part of the
wavefunction. However, in looking at the entanglement between spins, tracing out
over the momentum is implied.

\begin{figure}[htb]
\centerline{\psfig{figure=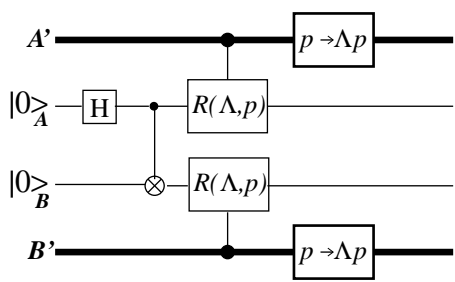,width=6cm}}
\bigskip
\caption{\label{circ}Circuit diagram for a Lorentz boost on a state
  with spins in a $|\phi^+ \rangle$ state. Lines representing momentum
  degrees of freedom are bold. }
\end{figure}

The wavefunction for two massive spin-1/2 particles can be written as \begin{equation}
|\Psi_{AA'BB'}\ra = \int\int \sum_{\lambda \sigma} g_{\lambda \sigma}({\bf
p},{\bf
  q})\, {|{\bf p} \lambda \ra}_{AA'} |{\bf q} \sigma \ra_{BB'}\, \widetilde{\rm d}{\bf p}\,
\widetilde{\rm d}{\bf q}\;, \end{equation} where $\widetilde{\rm d}{\bf p}$ and
$\widetilde{\rm d}{\bf q}$ are the Lorentz-invariant momentum integration measures
given by \begin{equation} \widetilde{\rm d}{\bf p} \equiv \frac{{\rm d}^3 {\bf
p}}{2 \sqrt{{\bf
      p}^2 + m^2}} \label{relat}
\end{equation}
and the functions $g_{\lambda \sigma}({\bf p},{\bf q})$ must
satisfy
\begin{equation}
\sum_{\lambda \sigma} \int\int |g_{\lambda \sigma}({\bf p},{\bf
  q})|^2 \widetilde{\rm d}{\bf p} \widetilde{\rm d}{\bf q} = 1\;.
\end{equation} To an observer in a frame Lorentz transformed by
  $\Lambda^{-1}$ the state $|\Psi_{AA'BB'}\ra$ appears to be
  transformed by $\Lambda \otimes \Lambda$.   Using
Eq. (\ref{lambdaonp}), and a change of variables for ${\bf p}$, ${\bf q}$,
$\lambda$ and $\sigma$, $g_{\lambda \sigma}({\bf
p},{\bf q})$ goes through the following transformation \begin{equation}
 g_{\lambda \sigma}({\bf p},{\bf q}) \rightarrow
 \sum_{\lambda' \sigma'} U^{(\Lambda^{-1}{\bf p})}_{\lambda,\lambda'}
U^{(\Lambda^{-1}{\bf q})}_{\sigma,\sigma'} g_{\lambda'
   \sigma'}(\Lambda^{-1} {\bf p},\Lambda^{-1} {\bf q})
\end{equation} where we defined
\begin{equation} U^{({\bf p})}_{\lambda,\lambda'} \equiv
D^{(\frac{1}{2})}_{\lambda,\lambda'} (R(\Lambda,{\bf p})) \end{equation} for compactness of
notation.  The Lorentz transformation can be viewed as a unitary operation,
$R(\Lambda,{\bf p})$, conditioned on ${\bf p}$ acting on the spin,
followed by a boost ${\bf p} \rightarrow \Lambda {\bf p}$ on the
momentum represented by the circuit  diagram in fig.~\ref{circ}.

By writing $|\Psi_{AA'BB'}\ra$ as a density matrix and tracing over the momentum
degrees of freedom, the entanglement between $A$ and $B$, (that is, between the
spin-degrees of freedom) can be obtained by calculating Wootters's
concurrence~\cite{Wootters98,fn1} \begin{equation}
C(\rho_{AB})=\max\{\lambda_1-\lambda_2-\lambda_3-\lambda_4,0\}\;,\end{equation} where
$\{\lambda_1,\lambda_2,\lambda_3,\lambda_4\}$ are the square roots of the
eigenvalues of the matrix $\rho_{AB}\tilde\rho_{AB}$, and $\tilde\rho_{AB}$ is the
``time-reversed" matrix~\cite{Wootters98} \begin{equation} \tilde\rho_{AB} =
(\sigma_y\otimes\sigma_y)\rho_{AB}^\star(\sigma_y\otimes\sigma_y)\;. \end{equation}

The first step in calculating the Lorentz transformed concurrence is to find an
explicit form for $U^{({\bf p})}_{\lambda,\lambda'}$. Since any Lorentz
transformation $\Lambda$ can be written as a rotation $R({\bf \Theta})$ followed
by a boost $L({\bf \xi})$ (see also Eq. (\ref{wigner})) it is clear that for a
pure rotation, $U^{({\bf p})}_{\lambda,\lambda'}$ does not depend on ${\bf p}$.
Hence, tracing over the momentum after a rotation will not change the concurrence.
Therefore, we can look only at pure boosts, and without loss of generality we may
choose boosts in the $z$-direction.  Writing the momentum 4-vector in polar
coordinates as \begin{equation} {\bf p} = [E,p \: \cos (\theta) \sin (\phi),p \: \sin (\theta)
\sin (\phi),p \: \cos (\phi)]\;, \end{equation} we obtain

\begin{equation} \label{matrix}
U^{({\bf p})} = \left[ \begin{array}{cc} \alpha & \beta e^{- i
      \theta} \\ - \beta e^{i \theta} & \alpha \end{array} \right] \;,
\end{equation}
where
\begin{eqnarray} \alpha & = & \sqrt{\frac{E + m}{E' + m}} \left[ \cosh \left(
\frac{\xi}{2} \right) + \frac{p \cos (\phi)}{(E + m)} \sinh \left( \frac{\xi}{2}
\right) \right] \;,\\
\beta & = & \frac{p \sin (\phi)}{\sqrt{(E + m)(E' + m)}} \sinh \left( \frac{\xi}{2}
  \right)\;,
\end{eqnarray}
and
\begin{equation}
E' =  E \cosh (\xi) + p \cos (\phi) \sinh (\xi).
\end{equation}
Here we use $\xi = |{\bf \xi}|$ as the rapidity of the boost in the
$z$ direction.

For momentum distributions in this letter we use a ``relativistic gaussian'' with
width $\sigma_r$
 \begin{equation} f({\bf p}) =\sqrt{\frac{1}{N(\sigma_r)}
 \exp{-(\frac{p^2}{2 \sigma_r^2})}}\;,
 \end{equation}
which differs from the standard gaussian only in the normalization $N(\sigma_r)$,
chosen in accordance with (\ref{relat}).

\begin{figure}[htb]
\centerline{\psfig{figure=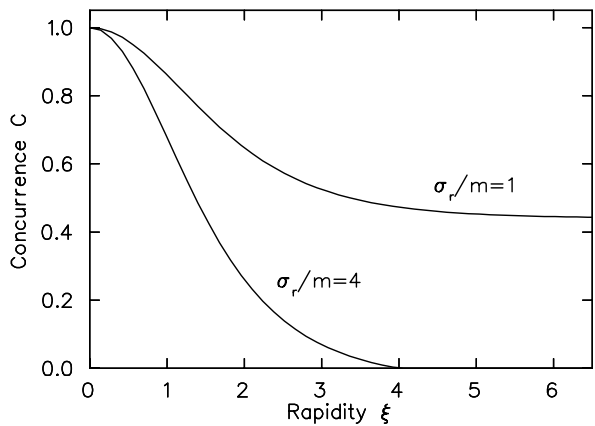,width=5.5cm,angle=90}}
\bigskip
\caption{\label{conc} Spin-concurrence as a function of rapidity, for an
initial Bell state with momenta in a product gaussian.  Data is shown for $\sigma_r / m = 1$ and
$\sigma_r / m = 4$.}
\end{figure}

For a spin Bell state $|\phi^+ \ra$ with momenta in a product gaussian, we have
 \begin{equation} \label{spinstate}
g_{\lambda \sigma}({\bf p},{\bf q}) = \frac{1}{\sqrt{2}} \delta_{\lambda
\sigma} f({\bf p}) f({\bf q}). \end{equation} Boosting this state, we move some of the
spin entanglement to the momentum. Tracing out the momentum from the
Lorentz-transformed density matrix destroys some
of the entanglement, and hence the concurrence in the moving frame
diminishes. The change in concurrence depends only on the ratio
$\sigma_r / m$ and $\xi$.  Figure \ref{conc} shows the concurrence vs.\ rapidity $\xi$,
for $\sigma_r / m = 1$ and $4$.  The decrease from the maximum value (the
concurrence is one for Bell states) documents the boost-induced decoherence of the
spin entanglement.

In the limit $\xi \rightarrow \infty$ (boost to the speed of light), the
concurrence saturates, i.e., it reaches a
constant value that depends on the mass of the particles and the shape of the
momentum distribution. In this particular example, it depends on the
ratio $\sigma_r / m$.  The saturation level decreases as $\sigma_r /
m$ increases until $\sigma_r / m \simeq 3.377$ when the saturation
level becomes zero.  Note that in the limit of ``pure"
momentum states (plane waves), the spins undergo local unitary
rotations  but entanglement
transfer does not occur, as was observed in \cite{czachor,alsingmil02}.
The reason for the saturation can be seen by examining (\ref{matrix})
in the limit $\xi\rightarrow\infty$

 \begin{eqnarray} \label{ablimit1}
\lim_{\xi \rightarrow \infty} \alpha & = & \sqrt{\frac{E + m}{2 (E + p
    \cos (\phi))}} \left( 1 + \frac{p \cos (\phi)}{E + m} \right) \\
\label{ablimit2} \lim_{\xi \rightarrow \infty} \beta & = & \frac{p \sin (\phi)}{\sqrt{2 (E +
  m) (E + p \cos (\phi))}}\;.
\end{eqnarray}
The parameter $\beta$ represents the amount of rotation due to the boost.  If
we maximize $\beta$ with respect to $\phi$ we obtain \begin{equation} \label{bmax} \beta
\rightarrow \frac{\frac{p}{m}}{1 + \sqrt{1 + \left( \frac{p}{m}
    \right)^2}}\;,
\end{equation} which is a monotonically increasing function of $p/m$.  For a particle of mass
$m$ and magnitude of momentum $p$, Eq.\ (\ref{bmax}) represents the maximal amount
of rotation due to a boost.  By increasing $\sigma_r / m$ in our
example, we effectivly increase this limit for $\beta$, and hence how much we can alter the
concurrence.  Note that since for large $p/m$ the rotation $\beta$ tends to one,
the Lorentz transformation is equivalent to a conditional spin flip in this regime.

If boosts can disentangle spins, can they transfer entanglement from the momentum
degrees of freedom to an unentangled spin wavefunction? Indeed this is possible.
One way to achieve this is to take any of the resulting states after boosting the
state in Eq.~(\ref{spinstate}) and apply the inverse boost to increase concurrence
to one.  Note that the increase in spin entanglement comes at the expense of a loss
of momentum entanglement, since the entanglement between all degrees of freedom
(spin and momentum) is constant under Lorentz transformations.

Simply reversing a previously applied Lorentz
transformation as in the last example is not a very satisfying way to
create entanglement.  Is there a way we could create an unentangled
state in the laboratory frame that would appear entangled to a moving
observer?  Consider the state
\begin{equation}
|\Psi_{AA'BB'}\ra = \frac{1}{\sqrt{2}} \left( |{\bf p},-{\bf p}
  \ra
 |\phi^+ \ra + |{\bf p}_\perp,-{\bf p}_\perp \ra
 |\phi^- \ra \right)
\end{equation}
where ${\bf p}$ and ${\bf p}_\perp$ are both in the $x$, $y$ plane, have the same
magnitude $p$, and are perpendicular. We could imagine such a state arising from a
particle decay where the products are restricted to movement in the $x$ or $y$ axes
with a conditional $\sigma_z$ gate on the perpendicular direction.  The reduced
density matrix $\rho_{AB}$ for this wavefunction is separable, and its concurrence
vanishes. However, taking the large $p$ limit in Eqs.~(\ref{ablimit1}) and
(\ref{ablimit2}) and choosing $\phi$ and $\theta$ appropriate for ${\bf p}$ and
${\bf p}_\perp$ in the $x$ and $y$ directions respectively, one can show that for a
large boost in the $z$ direction both $|\phi^+ \ra$ and $|\phi^- \ra$ are
transformed into the $|\psi^- \ra$ state and hence the spins are maximally
entangled in this reference frame.  In fact, the concurrence as a function of ${\bf
p}$ and $\xi$ is given by
\begin{equation}
C(\rho_{A B}) = \frac{p^2 (\cosh^2 (\xi) - 1)}{(\sqrt{1 + p^2} \cosh
  (\xi) + 1)^2}\;,
\end{equation}
when choosing $m = 1$.  Note that the concurrence is greater than zero whenever $p$
and $\xi$ are nonzero, and as $p$ and $\xi$ become large the concurrence tends to
one. So, if we restrict ourselves to spin measurements, an observer in the rest
frame of the decay particle cannot use entanglement as a resource (e.g., for
teleportation, super-dense coding, etc.) while the moving observer {\it can}. Such
a purification of spin entanglement is not always possible, however, and the
following theorem characterizes the limitations:

{\bf Theorem:} The entanglement {\it between} the spin and momentum parts of a pure
state wave function, $|\Psi_{AA'BB'}\ra$, must be non-zero to allow the spin
entanglement to increase under Lorentz transformations.

Proving the contrapositive, starting with a product state of the form
\begin{equation}
|\Psi_{AA'BB'}\ra = |\psi \ra_{A' B'} |\phi \ra_{A B}
\end{equation}
and applying boosts of the form $\Lambda \otimes \Lambda$ or even $\Lambda \otimes
\Lambda'$, we obtain
\begin{equation}
\label{thrmrho} \rho_{A B} = \sum_{i} p_i U_A^i \otimes V_B^i |\phi \ra \la \phi|
U_A^{i \dagger} \otimes V_B^{i \dagger}\;,
\end{equation}
where $U_A^i$ and $V_B^i$ are unitary operators and the sum $\sum_i p_i$ will be an
integral for certain states $|\psi \ra$. We can now plug Eq. (\ref{thrmrho})into
any entanglement monotone $E(\rho)$ and obtain the inequality
\begin{eqnarray}
\label{firstinequ}
  E(\rho_{A B}) & \leq & \sum_i p_i E(U_A^i \otimes V_B^i |\phi \ra)
  \\
  & = & \sum_i p_i E(|\phi \ra) \\
  & = & E (|\phi \ra)
\end{eqnarray}
where inequality (\ref{firstinequ}) comes from the definition of an entanglement
monotone~\cite{vidal98}.  Hence, the spin entanglement can only decrease after a
Lorentz transformation.  $\Box$

Note that this theorem does not hold if arbitrary unitary operations are
applied to a particle's spin and momentum degrees of freedom (for instance a swap
gate), but it does hold for the entire class of unitaries realized by Lorentz transformations.

We have investigated the properties of moving entangled pairs of massive particles.
Because Lorentz boosts entangle the spin and momentum degrees of freedom,
entanglement can be transferred between them. This is true for single
particles~\cite{peresetal02}, and we have shown here that it is true for pairs,
where the Lorentz boost affects the entanglement \textit{between} spins. Quite
generally, we can say that fully entangled spin states will (depending on the
initial momentum wavefunction) most likely decohere due to the mixing with momentum
degrees of freedom. We also note, however, that such mixing can \textit{purify}
spin entanglement if the momentum degrees are entangled with the spin. The physics
of creating entanglement between spins and between momenta is very different. Thus,
the possibility of entanglement transfer via Lorentz  boosts could conceivably, in
special situations, lead to simplified state preparation and purification
protocols.

We would like to thank Jonathan Dowling, and the members of the JPL Quantum
Computing Group, for useful discussions and encouragement. We also acknowledge
Daniel Terno for valuable suggestions. This work was carried out at the Jet
Propulsion Laboratory (California Institute of Technology) under a contract with
the National Aeronautics and Space Administration, with support from
the National Security Agency, the Advanced Research and Development
Activity, the Defense Advanced Research Projects Agency, the National
Reconnaissance Office, and the Office of Naval Research.

\end{multicols}

\begin{thebibliography}{99}
\bibitem{fn1}Because after the Lorentz transform the spins are in a mixed state,
entanglement measures based on pure states cannot be used.
\end{thebibliography}
\end{document}